\documentclass[aps,twocolumn,superscriptaddress,pra,showpacs]{revtex4-1}
\usepackage{amsmath} \usepackage{MnSymbol}

\usepackage{graphicx, color}

 \definecolor{BLACK}{gray}{0}
 \definecolor{WHITE}{gray}{1}
 \definecolor{RED}{rgb}{1,0,0}
 \definecolor{GREEN}{rgb}{0,.5,0}
 \definecolor{BLUE}{rgb}{0,0,1}
 \definecolor{CYAN}{cmyk}{1,0,0,0}
 \definecolor{MAGENTA}{cmyk}{0,1,0,0}
 \definecolor{YELLOW}{cmyk}{0,0,1,0}
 \definecolor{sblue}{rgb}{0.03,0.14,1.0}

\newcommand{\expect}[1]{\langle #1 \rangle}

\def\var{{\rm var \,}}

\usepackage[colorlinks,linktocpage,urlcolor=green]{hyperref}

\begin{document}

\title{Optimal Signal Recovery for Pulsed Balanced Detection} 

\author{Yannick A. de Icaza Astiz} 
\email[Corresponding author:]{yannick.deicaza@icfo.es}
\affiliation{ICFO-Institut de Ciencies
  Fotoniques, Mediterranean Technology Park, 08860 Castelldefels
  (Barcelona), Spain} 
\author{Vito Giovanni Lucivero}
\affiliation{ICFO-Institut de Ciencies
  Fotoniques, Mediterranean Technology Park, 08860 Castelldefels
  (Barcelona), Spain}  
\author{R. de J. Le\'{o}n-Montiel} 
\affiliation{ICFO-Institut de Ciencies
  Fotoniques, Mediterranean Technology Park, 08860 Castelldefels
  (Barcelona), Spain} 

\author{Morgan W. Mitchell} 
\affiliation{ICFO-Institut de Ciencies
  Fotoniques, Mediterranean Technology Park, 08860 Castelldefels
  (Barcelona), Spain} 
\affiliation{ICREA-Instituci\'{o} Catalana de
  Recerca i Estudis Avan\c{c}ats, 08015 Barcelona, Spain}

\date{\today}

\begin{abstract}
  We demonstrate a new tool for filtering technical and electronic
  noises from pulses of light, especially relevant for signal
  processing methods in quantum optics experiments as a means to
  achieve the shot-noise level and reduce strong technical noise by
  means of a pattern function. We provide the theory of this
  pattern-function filtering based on balance detection. Moreover, we
  implement an experimental demonstration where 10~dB of technical
  noise is filtered after balance detection. Such filter can readily
  be used for probing magnetic atomic ensembles in environments with
  strong technical noise.
\end{abstract}

\pacs{42.62.Eh, 42.50.Lc, 42.50.Dv,
  07.05.Kf}

\maketitle

\section{Introduction}
\label{sec:intro}

Balanced detection provides a unique tool for many physical,
biological and chemical applications. In particular, it has proven
useful for improving the coherent detection in telecommunication
systems \cite{Painchaud2009,Bach2005}, in the measurement of
polarization squeezing \cite{Loudon1987, Banaszek1997, Zhang1998,
  Predojevic2008, Agha2010}, for the detection of polarization states
of weak signals via homodyne detection \cite{Youn2005,Youn2012}, and
in the study of light-atom interactions
\cite{Kubasik2009}. Interestingly, balanced detection has proved to be
useful when performing highly sensitive magnetometry \cite{Sheng2013,
  Budker2007}, even at the shot-noise level, in the continuous-wave
\cite{Wolfgramm2010, Lucivero2014} and pulsed regimes
\cite{Koschorreck2010, Behbood2013}.

The detection of light pulses at the shot-noise level with low or
negligible noise contributions, namely from detection electronics
(electronic noise) and from intensity fluctuations (technical noise),
is of paramount importance in many quantum optics experiments. While
electronic noise can be overcome by making use of better electronic
equipment, technical noise requires special techniques to filter it,
such as balanced detection and spectral filtering.

Even though several schemes have been implemented to overcome these
noise sources \cite{Hansen2001, Chen2007, Windpassinger2009a}, an
optimal shot-noise signal recovery technique that can deal with both
technical and electronic noises, has not been presented yet. In this
paper, we provide a new tool based both on balanced detection and on
the precise calculation of a specific pattern function that allows the
optimal, shot-noise limited, signal recovery by digital filtering. To
demonstrate its efficiency, we implement pattern-function filtering in
the presence of strong technical and electronic noises. We demonstrate
that up to 10~dB of technical noise for the highest average power of
the beam, after balanced detection, can be removed from the signal.
This is especially relevant in the measurement of
polarization-rotation angles, where technical noise cannot be
completely removed by means of balanced detectors
\cite{Ruilova-Zavgorodniy2003}. Furthermore, we show that our scheme
outperforms the Wiener filter, a widely used method in signal
processing \cite{Vaseghi2000}.

The paper is organized as follows. In section \ref{sec:pattern} we
present the theoretical model of the proposed technique, in section
\ref{sec:experiment} we show the operation of this tool by designing
and implementing an experiment, where high amount of noise (technical
and electronic) is filtered. Finally in section \ref{sec:conclusions}
we present the conclusions.

\section{Theoretical model}
\label{sec:pattern}

To optimally recover a pulsed signal in a balanced detection scheme,
it is necessary to characterize the detector response, as well as the
``electronic'' and ``technical'' noise contributions
\cite{Bachor2004}.  We now introduce the theoretical framework of the
filtering technique and show how optimal pulsed signal recovery can be
achieved.

\subsection{Model for a balanced detector}
\label{sec:realbalanced}

\begin{figure}[t]
  \centering 
  \includegraphics[width=0.48 \textwidth]{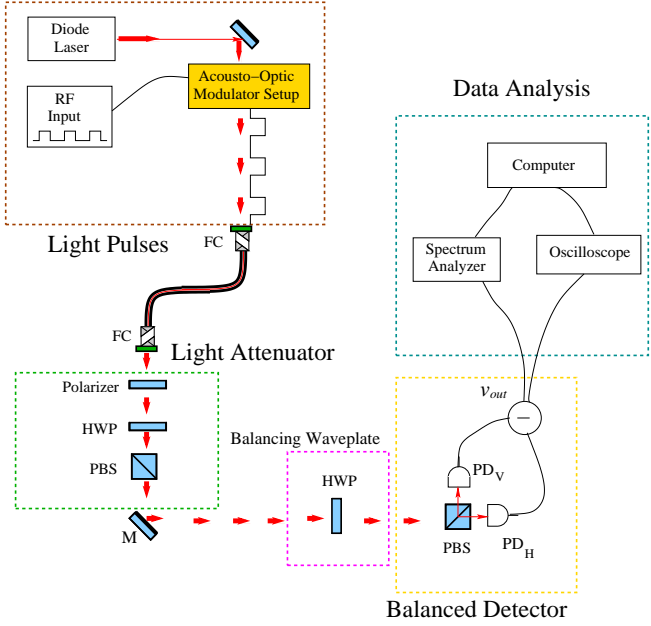}
\caption{Scheme of the experimental setup. M, mirror, FC, fiber
  coupling, HWP, half-wave plate. See text for details.}
  \label{fig:expsetup}
\end{figure}

To model a balanced detector, see Fig. \ref{fig:expsetup}, we assume
that it consists of 1) a polarizing beam splitter (PBS), which splits
the $H$ and $V$ polarization components to 
two different detectors 2)
the two detectors PD$_H$ and PD$_V$, whose output currents are
directly subtracted, and 3) a linear amplifier

Because the amplification is linear and stationary, we can describe
the response of the detector by impulse response functions $h(\tau)$.
If the photon flux at detector {$X$} is {$\phi_X(t)$}, the electronic
output can be defined as

\begin{equation}
  \label{eq:impulse1}
  v_{\rm out}(t) \equiv h_H *\phi_H  + h_V*\phi_V + v_{N}(t),   
\end{equation} where $v_N$ is the electronic noise of the photodiodes, including amplification. Here, $h*\phi$ stands for the convolution of $h$ and $\phi$, i.e., $(h * \phi)(t) \equiv \int_{-\infty}^{\infty} h(t-\tau) \phi(\tau)d\tau$. For clarity, the time dependence will be suppressed when possible. It is convenient to introduce the following notation: $\phi_S \equiv
\phi_H + \phi_V$, $\phi_D \equiv \phi_H - \phi_V$, $h_S \equiv h_H +
h_V$ and $h_D \equiv h_H - h_V$. Using these new variables, Eq. \eqref{eq:impulse1} takes the form

\begin{equation}
  \label{eq:impulse2}
  v_{\rm out}(t) = \frac{1}{2}\left( {h}_S * {\phi}_S +
    {h}_D * {\phi}_D \right) + v_{N}(t).
\end{equation} 

From this signal, we are interested in recovering the differential
photon number $S \equiv \int_\mathcal{T} \phi_H(t)dt -
\int_\mathcal{T} \phi_V(t)dt$, where $\mathcal{T}$ is
  the time interval of the desired pulse, with minimal uncertainty.
More precisely, we want to find an estimator $\hat{S}[v_{\rm
  out}(t)]$, that is unbiased $\langle\hat{S}\rangle =
\langle{S}\rangle$, and has minimal variance $\var (\hat{S})$.

\subsection{Signal recovery estimator}
\label{sec:estimation1}

In order to make $\hat{S}$ unbiased, we realize that it must linearly
depend on $v_{\rm out}$. This because $S$ and $v_{\rm out}$ are linear
in both $\phi_H$ and $\phi_V$. Therefore, the estimator must have the
form
\begin{equation}
  \label{eq:raw0}
  \hat{S} = \int_{-\infty}^{\infty}  v_{\rm out}(t) \gamma(t) dt.
\end{equation}

In Eq. \eqref{eq:raw0}, $\gamma(t)$ refers to as \emph{pattern
  function}, which describes the most general linear estimator. In
this work, we will consider three cases: 1) a raw estimator,
$\gamma(t)=1$ for $t \in \mathcal{T}$ and 0 otherwise; 2) a
Wiener estimator, which makes use of a Wiener-filter-like pattern
function, $\gamma(t)=w(t)$, where $w(t)$ represents the Wiener filter
in the time domain \cite{Vaseghi2000}, and 3) a model-based pattern
function estimator $\gamma(t)=g(t)$.  Notice that both $w(t)$
  and $g(t)$ are defined in $(-\infty,\infty)$, allowing to properly
  choose a desired pulse. In what follows, we explicitly show how to
calculate the model-based pattern function estimator $g(t)$.

\subsection{Conditions of the pattern function}
\label{sec:conditions}

We assume that $\phi_S,\phi_D$ have known averages (over many pulses)
$\bar{\phi}_S(t),\bar{\phi}_D(t)$, and similarly the response
functions $h_S(\tau), h_D(\tau)$ have averages $\bar{h}_S(\tau),
\bar{h}_D(\tau)$.  Then the average of the electronic output reads as
\begin{equation}
  \label{eq:modsignal2}
  \bar{v}_{out}(t)  = \frac{\overline{h}_S * 
    \overline{\phi}_S + \overline{h}_D * \overline{\phi}_D}{2},
\end{equation}
and $\expect{\hat{S}} = \int_{-\infty}^{\infty} dt\, g(t)
\left( \bar{h}_S * \bar{\phi}_S + \bar{h}_D * \bar{\phi}_D
\right)/2$. In writing Eq. \eqref{eq:modsignal2}, we have assumed
  that the noise sources are uncorrelated.
 
From this we observe that if a balanced optical signal is introduced,
i.e. $\bar{\phi}_D = 0$, the mean electronic signal $\bar{v}_{out}(t)$
is entirely due to $\overline{h}_S * \overline{\phi}_S$.  In order
that $\hat{S}$ correctly detects this null signal, $g(t)$ must be
orthogonal to $\overline{h}_S * \overline{\phi}_S$, i.e.
  
\begin{equation}
\label{eq:orthogonality}
\int_{-\infty}^{\infty}  g(t)\cdot \left( \overline{h}_S * \overline{\phi}_S \right)(t) dt = 0.
\end{equation}

Our second condition derives from
\begin{equation}
  \label{eq:calibration0}
  \int_{-\infty}^{\infty}  g(t) \cdot \left( \overline{h}_D * \overline{\phi}_D \right)(t) dt=
  \int_\mathcal{T}  \overline{\phi}_D(t)  dt,
\end{equation} which is in effect a calibration
condition: the right-hand side is a uniform-weight integral of $\overline{\phi}_D$, while the left-hand side is a non-uniform-weight integral, giving preference to some parts of the signal.  If the total weights are the same, the
above gives $\expect{\hat{S}} = \expect{S}$. We note that this
condition is not very restrictive.  For example, given
$\bar{h},\bar{\phi}$, and given $g(t)$ up to a normalization, the
equation simply specifies the normalization of $g(t)$. 

Notice that the condition given by Eq. \eqref{eq:calibration0} may
still be somewhat ambiguous. If we want this to apply for all possible
shapes $\bar{\phi}_D(t)$, it would imply $g(t) =$ const., and would
make the whole exercise trivial. Instead, we make the physically
reasonably assumption that the input pulse, with shape $\bar{\phi}_S$
is uniformly rotated to give $\bar{\phi}_H(t)$, $\bar{\phi}_V(t)
\propto \bar{\phi}_S$. Similarly, it follows that $ \bar{\phi}_D(t)
\propto \bar{\phi}_S$. We note that this assumption is not strictly
obeyed in our experiment and is a matter of mathematical convenience:
a path difference from the PBS to the two detectors will introduce an
arrival-time difference giving rise to opposite-polarity features at
the start and end of the pulse, as seen in Fig. \ref{fig:Restech}(a).
A delay in the corresponding response functions $h$ is, however,
equivalent, and we opt to absorb all path delays into the response
functions.  In our experiment the path difference is $\approx5~\mathrm{cm}$,
implying a time difference of less than 0.2~ns, much below the
smallest features in Fig. \ref{fig:Restech}(a).  Absorbing the
constant of proportionality into $g(t)$, we find

\begin{equation}
  \label{eq:calibration}
  \int_{-\infty}^{\infty}  g(t) \cdot \left( \overline{h}_D * \overline{\phi}_S \right)(t) dt=
  \int_\mathcal{T}  \overline{\phi}_S(t)  dt,  
\end{equation} which is our calibration condition.

\subsection{Noise model}
\label{sec:RNM}

We consider two kinds of technical noise: fluctuating detector
response and fluctuating input pulses.  We write the response
functions in the form $h_X = \bar{h}_X + \delta h_X$, for a given
detector $X$, where the fluctuating term $\delta h_X$ is a stochastic
variable. Similarly, we write ${\phi}_Y = \bar{\phi}_Y + \delta
\phi_Y$, where $Y$ is $H, V, S$ or $D$.  By substituting the
corresponding fluctuating response functions into
Eq. \eqref{eq:impulse2}, the electronic output signal becomes

\begin{eqnarray}
  \label{eq:techmodel1MWM}
    v_{out}(t) & = &\frac{1}{2}\left( \overline{h}_S * \overline{\phi}_S   +    \overline{h}_D * \overline{\phi}_D \right)
    +v_{N}(t) 
    \nonumber \\ & & + 
 \frac{1}{2}\left( \delta {h}_S * \overline{\phi}_S +
      \delta {h}_D * \overline{\phi}_D \right) 
         \nonumber \\ & & + 
 \frac{1}{2}\left(  \overline{h}_S * \delta {\phi}_S +
       \overline{h}_D *  \delta{\phi}_D \right)  
                + 
O(\delta h \, \delta \phi)  
     \\ & \approx & \frac{1}{2}\left( \overline{h}_S * \overline{\phi}_S   +    \overline{h}_D * \overline{\phi}_D \right)
    +v_{N}(t) + v_{T}(t),
\end{eqnarray}
where $v_{T}(t) \equiv \frac{1}{2}( \delta {h}_S * \overline{\phi}_S +
\delta {h}_D * \overline{\phi}_D + \overline{h}_S * \delta {\phi}_S +
\overline{h}_D * \delta{\phi}_D ) $ is the summed technical noise from
both $\delta h$ and $\delta \phi$ sources.  We note that the optical
technical noise, in contrast to optical quantum noise, scales as
$\var(\delta \phi) \propto \bar{\phi}^2$, so that $\var(v_T) \propto
\bar{\phi}^2$.  In passing to the last line we neglect terms $O(\delta
h \, \delta \phi)$ on the assumption $\delta h \ll \bar{h}$, $\delta
\phi \ll \bar{\phi}$.  We further assume that $v_{N}$ and $v_{T}$ are
uncorrelated.

We find the variance of the model-based
estimator, $N_\sigma \equiv \var(\hat{S}_{\mathrm{opt}})$, is
\begin{equation}
  \label{eq:NPvariace}
  N_\sigma= \left\langle \left| \int_{-\infty}^{\infty}  g(t) v_T(t) dt \right|^2\right\rangle   + \left< \left| \int_{-\infty}^{\infty}  g(t) v_{N}(t) dt \right|^2 \right>,
\end{equation} with the first term describing technical noise, and the second one electronic noise. 

To compare against noise measurements, we transform
Eq. \eqref{eq:NPvariace} to the frequency domain.
Using Parseval's theorem, see Eq. (\ref{eq:simp1MWM}), we can write the 
noise power as

\begin{equation}
  \label{eq:NPvariace2}
  \begin{split}
  N_\sigma= &  \int_{-\infty}^{\infty}  |G(\omega)|^2   \langle|V_T(\omega)|^2 +   | V_{N}(\omega)|^2 \rangle  d\omega.
  \end{split}
\end{equation}

Our goal is now to find the
$G(\omega)$ that minimizes  $N_\sigma$
satisfying the conditions in Eqs. \eqref{eq:orthogonality} and
\eqref{eq:calibration}, which in the frequency space are

\begin{equation}
  \label{eq:Orthogonality}
I_{\rm or} \equiv \int_{-\infty}^{\infty} d\omega \, G^*(\omega) \overline{H}_S(\omega) \overline{\Phi}_S(\omega) = 0,
\end{equation}

\begin{equation}
  \label{eq:Calibration}
  I_{\rm cal} \equiv \int_{-\infty}^{\infty} d\omega \, G^*(\omega)  \overline{H}_D(\omega)  \overline{\Phi}_S (\omega) = \overline{\Phi}_S(0).
\end{equation}

The specific form of the solution is given in Appendix
\ref{sec:pattesolution}.

\section{Experiment}
\label{sec:experiment}

\begin{figure}[t!]
  \centering
    \includegraphics[width=6cm]{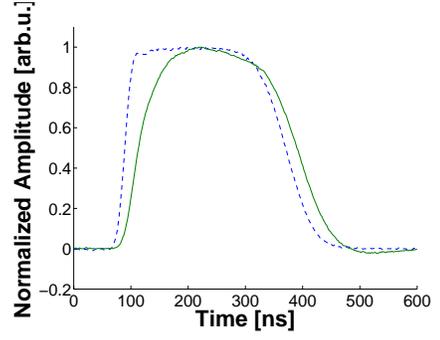}
    \caption{Average pulse shapes of the original pulse $p(t)$ at
      150~MHz (blue dashed line) and the amplified one $p'(t)$ at
      5~MHz (green solid line). For the sake of comparison, both
      pulses are normalized.}
  \label{fig:E}
\end{figure}

\begin{figure}[h!]
  \centering
  \begin{tabular}{c}
    \includegraphics[width=9cm]{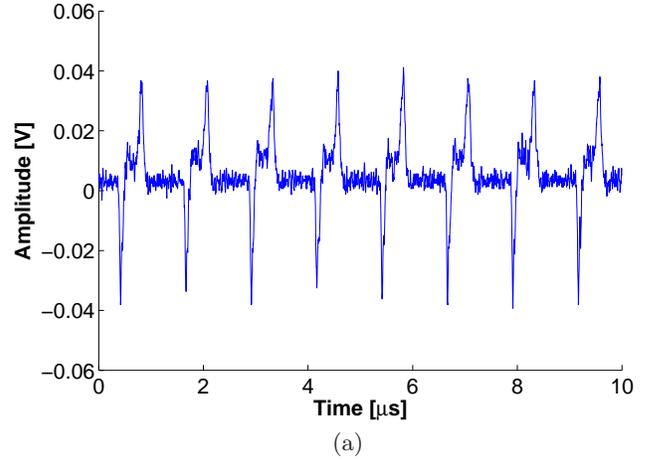}\\
    (a)\\
\includegraphics[width=9cm]{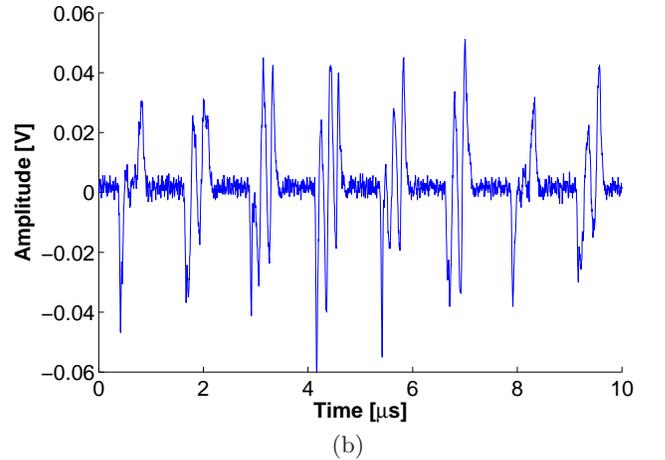}\\
(b)\\
  \end{tabular}
  \caption{Example of pulses seen by the balanced detector (a) without
    technical noise, and (b) with technical noise introduced.}
  \label{fig:Restech}
\end{figure}

\subsection{Pulse detection and detector characterization}

In our experimental setup, pulsed signals are produced using an
external cavity diode laser at 795~nm (Toptica DL100), modulated by
two acousto-optic modulators (AOMs) in series. We have used two AOMs
to prevent a shift in the optical frequency of the pulses, and also to
ensure a high extinction ratio $(r_e> 10^{7})$.

Balanced detection is performed by using a Thorlabs PDB150A detector
\cite{Thorlabs2007} that contains two matched photodiodes wired
back-to-back for direct current subtraction, amplified by a
switchable-gain transimpedance amplifier.  We use the gain settings
$10^3$~V/A and $10^5$~V/A, with nominal bandwidths of 150~MHz and
5~MHz, respectively. Figure \ref{fig:E}(a) shows the average pulse
shapes $p(t)$ and $p'(t)$, observed with bandwidth settings 150~MHz
and 5~MHz, respectively. These shapes are obtained by blocking one
detector and averaging over 1000 pulse traces (280~ns width).

In this way, to determine the {impulse response functions} $h_H(t)$,
$h_V(t)$ of the photodiodes PD$_H$ and PD$_V$, respectively, we first
assume the form

\begin{equation}
  \label{eq:tr2}
  h_X(t)=\frac{e^{-t/{\tau_\mathrm{TIA}}}-e^{-{t}/{\tau_X}}}{\tau_\mathrm{TIA}-\tau_X},
\end{equation} where $X \in \{H,V\}$ indicates the photodiode.  This describes a single-pole filter with time constant $\tau_X$ for  the photodiode \cite{Ezaki2006, Hamamatsu2012} followed by a single-pole  filter with time-constant $\tau_{\mathrm{TIA}}$ for the transimpedance amplifier. We  choose the
parameters $\tau_\mathrm{TIA},\tau_X$ by a least-squares fit of
\begin{equation}
  \label{eq:vq2}
  \tilde{p}'(t)\equiv \int_{-\infty}^{\infty} p(\tau) h_X(t-\tau)d\tau. 
\end{equation} to the measured traces $p'(t)$ \cite{Saleh2007}.

As seen in Fig. \ref{fig:Restech}(a), a small difference in the speeds
of the two detectors leads to electronic pulses with a negative
leading edge and a positive trailing edge, even when the optical
signal is balanced, i.e. even when the average electronic output is
zero.

\subsection{Producing technical noise in a controlled manner}
\label{sec:tech}

\begin{figure}[t!]
  \centering
\includegraphics[width=8.8cm]{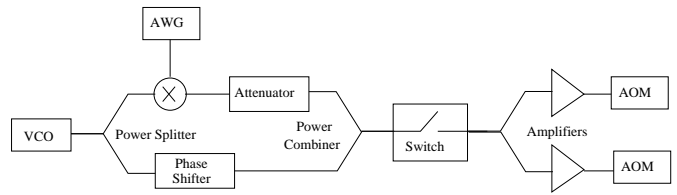}
\caption{Scheme of the electronic circuit used to introduce technical
  noise into the AOMs. See text for details.}
  \label{fig:techcircuit}
\end{figure}

\begin{figure}[h!]
  \centering 
\includegraphics[width=0.48
 \textwidth]{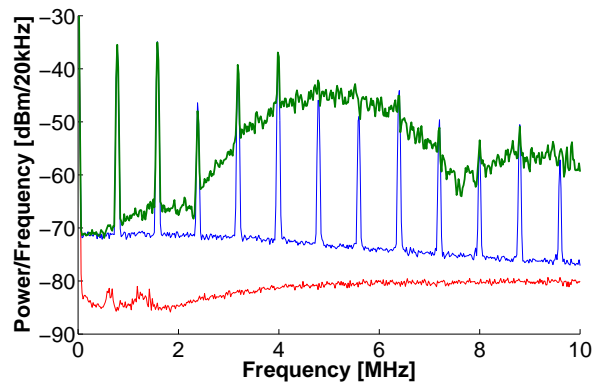}
\caption{Illustration of noise contributions in the power spectrum of
  a train of pulses.  Thin red curve shows the electronic noise of the
  detector, i.e., with no optical signal introduced.  Blue medium
  curve shows power spectrum with no introduced technical noise.  This
  shows narrow peaks at harmonics of the pulse repetition frequency
  rising from a shot-noise background.  The roll-off in signal
  strength is due to the 5 MHz bandwidth of the detector.  Thick green
  curve shows power spectrum with an introduced technical noise with
  central frequency of 5 MHz and FWHM bandwidth of 1 MHz.}
  \label{fig:G}
\end{figure}

In order to prove that it is possible to remove technical noise, first
we need to produce it in a controlled manner. To this end, we
introduce technical noise in our system perturbing the main frequency
of the AOMs using the circuit described in Fig. \ref{fig:techcircuit}.
 The main frequency is produced by a voltage
controlled oscillator (VCO) set to 80~MHz. Then, it is split with a
power splitter, one of the arms is mixed with a signal from an
arbitrary wave generator (AWG) and attenuated, whereas in the other
arm the signal is passed by a phase shifter. Finally, both signals are
put back together with a power combiner. In this way, we have a main
frequency of 80~MHz and sidebands at the frequency of the signal
introduced with the AWG. We can then program the AWG with technical
noise for a particular frequency and bandwidth, as illustrated in
Fig. \ref{fig:G}.

In our setup, we have fixed the parameters of the circuit and the AWG for
generating about 10~dB of technical noise for an optical power of
$400~\mu W$ with a duty cycle of the pulses of $1/3$.

\begin{figure}[h!]
 \centering
 \includegraphics[width=0.48
 \textwidth]{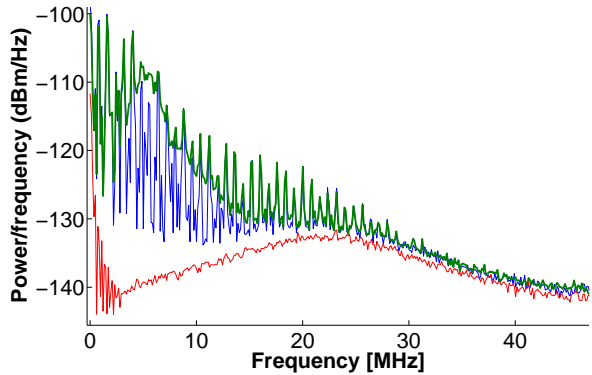}
 \caption{Power spectral density from a train of 800 pulses,
   considering three cases: signal without technical noise (blue
   line), signal with technical noise (bold green line), and
   electronic noise (red thin line). }
  \label{fig:Res1}
\end{figure}

\subsection{Calculating the optimal pattern function for different 
optical powers}

To measure the noise spectra upon which the pattern function will be
based, we use an oscilloscope (Lecroy Wavejet-324), rather than a
spectrum analyzer.  This allows us to use the same instrument for
noise characterization and optimization as we will later use to
acquire signals to process by digital filtering.

We collect $5 \times 10^5$ samples in a 1000~$\mu \text{s}$ acquisition
time containing a total of 800 pulses $\sim$400~ns duration, with a
duty cycle of 1/3. For this train of pulses we compute the power
spectral density (PSD) for three cases: 1) signal without added
technical noise, 2) signal with added technical noise, and 3)
the electronic noise. Figure \ref{fig:Res1} shows an example of PSD
calculated for these cases.  From these PSDs we can then extract the
parameters necessary for computing the optimal pattern function,
namely electronic background, technical noise power and shot-noise
power. Using these parameters, and following the method explained in
section \ref{sec:pattern}, we have calculated the optimal pattern
function $g(t)$ for different average powers of the beam, from 0 to
400~$\mu$W in steps of 20~$\mu$W.

\begin{figure}[h!]
  \centering
\includegraphics[width=9cm]{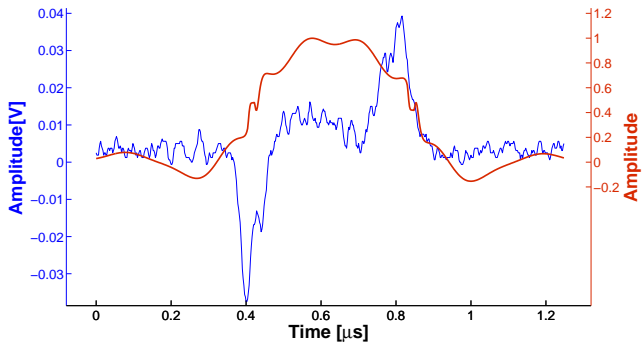}
\caption{Example of cutting of the pulses (blue thin line) and the
  corresponding pattern function (red thick line). }
  \label{fig:Res2}
\end{figure}

\subsection{Shot-noise limited detection with pulses and measurement
  of the technical noise with pulses}

Because the pulses are non-overlapping, as seen in
Fig. \ref{fig:Restech}, we can isolate any single pulse by keeping
only the signal in a finite window containing the pulse, to get a
waveform as illustrated in Fig. \ref{fig:Res2}.  Also shown there is
the optimal pattern function.  This illustrates some qualitative
features of the optimal pattern function, which is 1) orthogonal to
the residual common-mode signal $h_S * \phi_S$, which first goes
negative and then positive, 2) well overlapped with the
differential-mode signal $h_D * \phi_D$, which is positive, and 3)
smooth with some ringing, to suppress both high-frequency and
low-frequency noise.

\newcommand{\SRaw}{\hat{S}_{\rm raw}} 

\newcommand{\SWie}{\hat{S}_{\rm W}}

\newcommand{\SOpt}{\hat{S}_{\rm opt}}

For each pulse we compute the estimators $\SRaw$, $\SWie$ and $\SOpt$,
using pattern functions $\gamma(t)=1$ (raw estimator),
$\gamma(t)=w(t)$ (Wiener estimator) and $\gamma(t)=g(t)$ (optimal
model-based estimator), respectively. The Wiener filter $w(t)$ can be
defined as the Fourier transform of the frequency domain
representation of the Wiener filter $W(\omega)$, given by the
  ratio of the cross-power spectrum of the noisy signal with the
  desired signal over the power spectrum of the noisy signal
  \cite{Vaseghi2000}. For more details see the appendix
  \ref{sec:wienerest}.

\subsubsection{Shot-noise limited detection with pulses}

We first show that the system is shot-noise limited in the absence of
added technical noise.  For this, we compute the variance of $\SRaw$,
this variance is a noise estimation, computed from a pulse
train without technical noise, as a function of optical power $P$.  We
fit the resulting variances with the quadratic $\var(\SRaw)=A+BP+C
P^2$, and obtain $A=4.5\pm 0.3\times 10^{-20} J^2$, $B=2.4\pm 0.1
\times 10^{-22} J^2/\mu W$ and $C=6.7\pm0.6\times 10^{-26} J^2/\mu
W^2$. The data and fit are shown in Fig. \ref{fig:ResFit1}(a), and
clearly show a linear dependence on $P$, a hallmark of shot-noise
limited performance.

\subsubsection{Measuring technical noise with pulses}

Now, we proceed as before with the exception that in this case we
introduce technical noise to the signal. We obtain the following
fitting parameters: $A=4.5\pm 0.3\times 10^{-20} J^2$, $B=1.9\pm 0.1
\times 10^{-22} J^2/\mu W$ and $C=4.12\pm 0.05\times 10^{-24}
J^2/\mu W^2$.

\begin{figure}[h!]
  \centering 
\begin{tabular}{c}
\includegraphics[width=8.5cm]{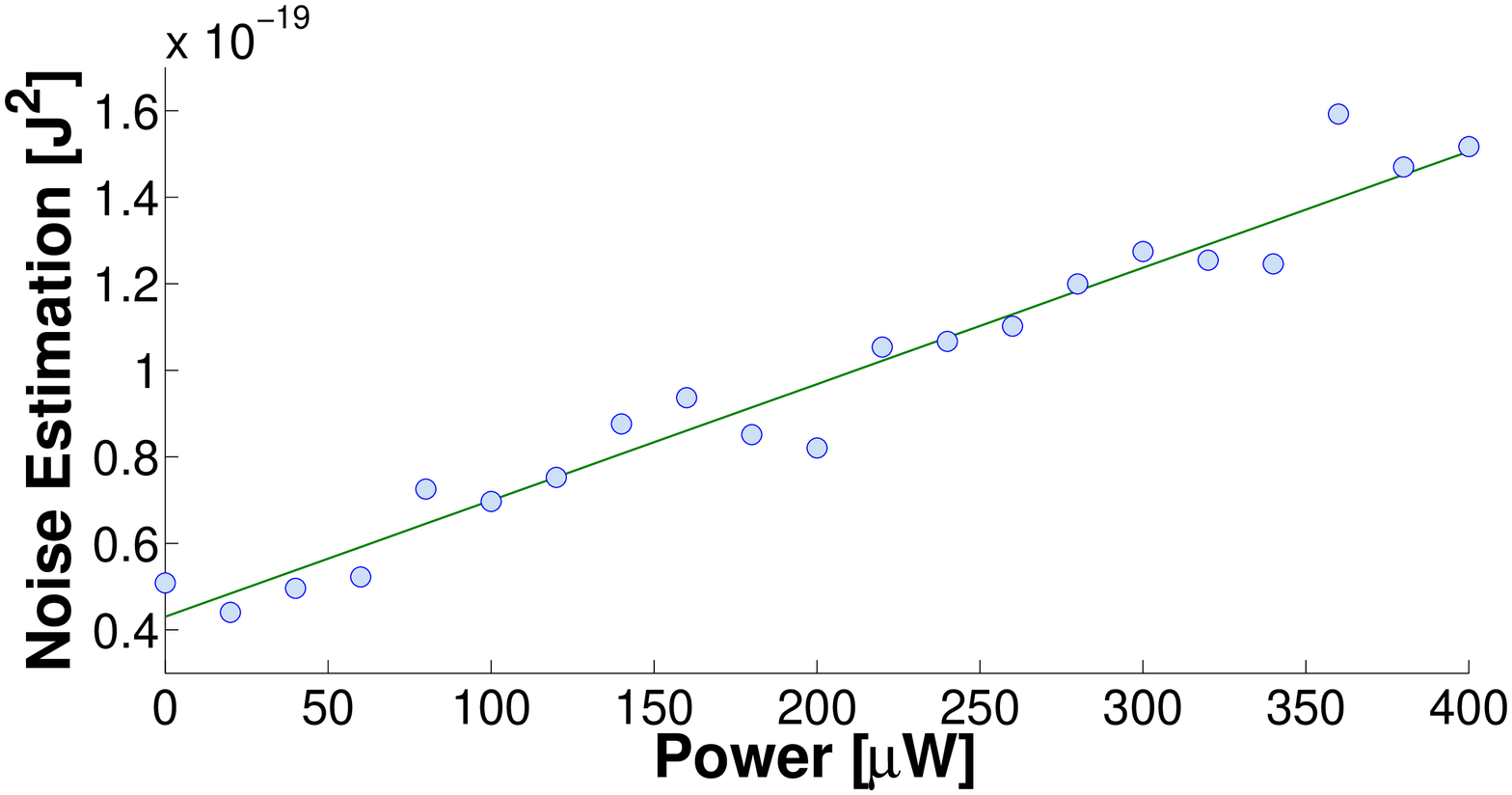}\\
(a)\\
\includegraphics[width=8.5cm]{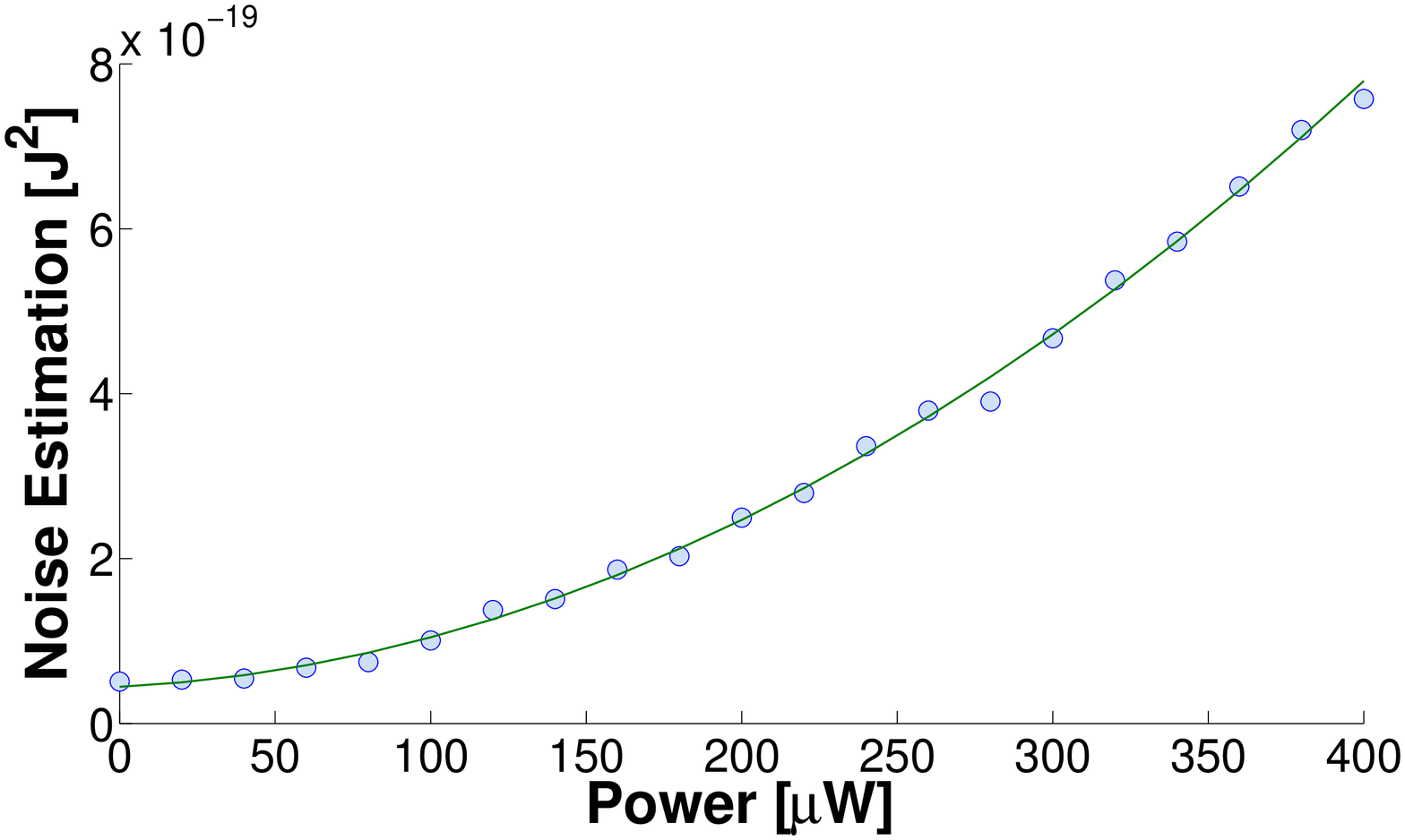}\\
(b)\\
\end{tabular}
\caption{Computed noise estimation as a function
  of the optical signal power (a) without and (b) with technical noise
  introduced. Circles: experimental data, solid line: quadratic fit. }
  \label{fig:ResFit1}
\end{figure}

We observe from Fig. \ref{fig:ResFit1}(b) that the noise estimation
for the data that has technical noise exhibits a clearly quadratic
trend, in contrast to the linear behavior where no technical noise is
introduced.  The results shown in Figs. \ref{fig:ResFit1}(a) and
\ref{fig:ResFit1}(b) prove that, with our designed system, it is
possible to introduce technical noise in a controlled way.

\subsection{Filtering 10~dB of technical noise using an optimal
  pattern function}
\label{sec:filtering10}

To illustrate the performance of our technique when filtering
technical noise, we introduce a high amount of noise ---about $60$ dB
above the shot noise level at the maximum optical power--- to the
light pulses produced by the AOMs. After balancing a maximum of 10~dB
remains in the electronic output, which is then filtered by means of
the optimal pattern function technique.

\begin{figure}[h!]
  \centering 
  \includegraphics[width=9cm]{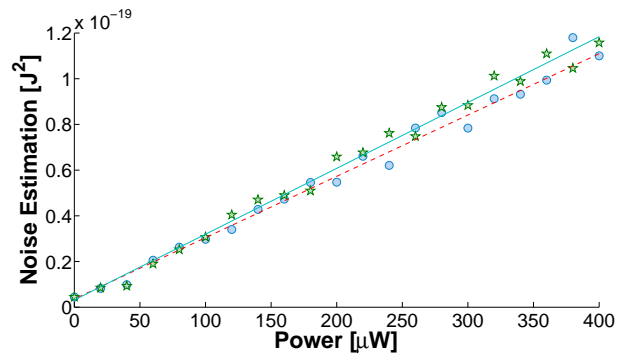}
  \caption{Computed noise estimation using the
    optimal pattern estimator as a function of the optical power for
    shot-noise limited pulses (blue circles) and pulses with technical
    noise (green stars). Their corresponding quadratic fits are shown
    in red dashed and cyan lines, respectively. }
  \label{fig:Res3}
\end{figure}

We have verified the correct noise filtering by comparing the results
with shot-noise limited pulses. For this purpose, we compute
$\var(\SOpt)$, the variance of the optimal estimator for each power,
and for each data set, the shot-noise limited and the noisy
one. Figure \ref{fig:Res3} shows the computed noise estimation as
function of the optical power for both. Notice that the two noise
estimations are linear with the optical
power. Moreover, we observe that both curves agree at $\sim 91\pm
5\%$, using the ratio of the slopes, which allows us to
conclude that, by using this technique, we can retrieve shot-noise
limited pulses from signals bearing high amount of technical noise.

\subsection{Optimal estimation of the polarization-rotation angle.}
\label{sec:phi}

The experimental setup that we have implemented, see
Fig. \ref{fig:expsetup}, can perform also as a pulsed signal
polarimeter. For instance, it is possible to determine a small
  polarization-rotation angle $\varphi$ from a $45^{\circ}$ linear polarized
  light pulse. Along these lines, we make use of three estimators
$\SRaw, \SWie$ and $\SOpt$ to determine the amount of noise on the
estimation of the polarization-rotation angle. From the obtained
results, we show that the model-based estimator outperforms the other
two.

We proceed to calculate the noise on the polarization-rotation angle
$\varphi$ estimation, for this determination we calculate the variance
of $\varphi$. We notice that the Taylor approximation of the variance
of $\hat{S}(\varphi)$ is

\begin{equation}
  \label{eq:phi2}
  \var(\hat{S})\approx\left(\frac{d \hat{S}}{d \varphi}\right)^2 \var (\varphi).
\end{equation} For small angles $\varphi$, the function $\hat{S}(\varphi)$ is approximately linear on $\varphi$, so the contribution from higher order terms can be disregarded.

Therefore, the noise on the angle estimation is

\begin{equation}
  \label{eq:angle1}
  \var(\varphi)= \frac{\var( \hat{S} )}{\left(\frac{d {\hat{S}}}{d\varphi}\right)^2}.
\end{equation}

\begin{figure}[h!]
  \centering
  \includegraphics[width=9.1cm]{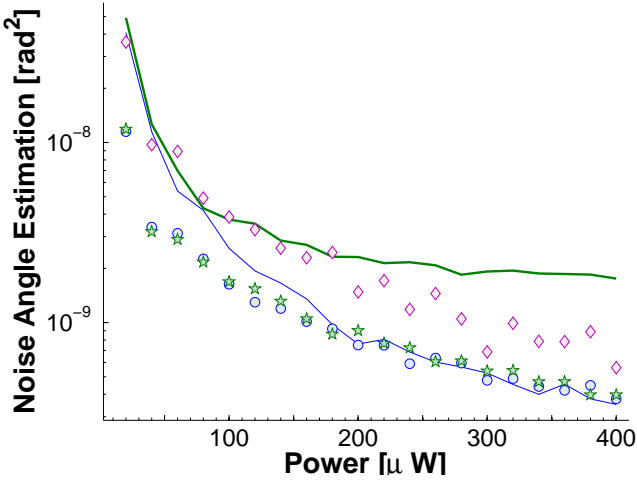}
\caption{Noise angle estimation as a function on the optical power.
  Raw estimators with technical noise (green bold line) and without
  (blue medium line). Wiener estimator with technical noise (pink
  diamonds). Model-based estimators with technical noise (green stars)
  and without (blue circles).  For the sake of visualization the
  results are presented in a semi-log graph. }
  \label{fig:Resangle}
\end{figure}

We can then compute this expression using the three before mentioned
estimators. For such task we use the experimental data together with
an analytical approximation of the derivative, that takes as input the
measured data. Figure \ref{fig:Resangle} depicts the noise angle
estimation, showing that the optimal pattern function performs better
than the other estimators when eliminating the technical noise and
reducing the electronic noise. In particular, the based-model
estimator surpasses the Wiener estimator, which is a widely used
method in signal processing \cite{Vaseghi2000}.

\section{Conclusions}
\label{sec:conclusions}

We have studied in theory and with an experimental demonstration, the
optimal recovery of light pulses via balanced detection. We developed
a theoretical model for a balanced detector and the noise related to
the detection of optical pulses. We minimized the technical and
electronic noise contributions obtaining the optimal (model-based)
pattern function. We designed and implemented an experimental setup to
test the introduced theoretical model. In this experimental setup, we
produced technical noise in a controlled way, and retrieved shot-noise
limited signals from signals bearing about 10~dB of technical noise
after balanced detection. Finally, we compare against na\"{i}ve and
Wiener filter estimation for measuring rotation angles, and confirm
superior performance of the model-based estimator.  

The results presented here might lead to a better
polarization-rotation angle estimations when using pulses leading to
probe magnetic atomic ensembles in environments with technical noise
\cite{Koschorreck2010, SewellPRL2012}. This possibility is
especially attractive for balanced detection of sub-shot-noise pulses
\cite{Predojevic2008, Wolfgramm2010}, for which the acceptable noise
levels are still lower.

\appendix

\section{Parseval}
\label{sec:Parseval}
We note the inner-product form of Parseval's theorem
\begin{equation}
  \label{eq:innerparseval}
  \int_{-\infty}^{\infty} G^*(t) x(t) dt= \int_{-\infty}^{\infty} G^*(\omega)X(\omega) d\omega,
\end{equation} where the functions $G(\omega), X(\omega)$ are the Fourier transforms of $g(t),x(t)$, respectively. For any stationary random variable $x(t)$,
$\expect{X(\omega)X(\omega')} = \delta(\omega-\omega')$ (if this were
not the case, there would be a phase relation between different
frequency components, which contradicts the assumption of stationarity). From this, it follows that
\begin{equation}
  \label{eq:simp1MWM}
  \left\langle \left| \int_{-\infty}^{\infty}  g(t) x(t) dt\right|^2\right\rangle =  \int_{-\infty}^{\infty} |G(\omega)|^2\langle |X(\omega)|^2\rangle d \omega.
\end{equation}

\section{Formal derivation of the pattern function}
\label{sec:pattesolution}

We will minimize the noise power $N_\sigma$ (see
Eq. \eqref{eq:NPvariace2}) with respect to the pattern function
$G(\omega)$ using the two conditions (see Eq. \eqref{eq:Orthogonality}
and Eq. \eqref{eq:Calibration}). We solve this by the method of
Lagrange multipliers.  For this, we write

\begin{equation}
  \label{eq:Lagrange1}
  L(G,\lambda_1,\lambda_2) = N_\sigma + \lambda_1 (I_{\rm or}-0) +
  \lambda_2 (I_{\rm cal}-\overline{\Phi}_S({\omega=0}) ), 
\end{equation}and then solve the equations

\begin{eqnarray}
  \label{eq:partials}
\partial_{G^*} L &=& 0, \nonumber \\
 \partial_{\lambda_1} L & =&  0, \nonumber \\
 \partial_{\lambda_2} L & =&  0.
\end{eqnarray}

The first equation reads

\begin{eqnarray}
  \partial_{G^*} L &=& G(\omega)  \langle|V_T(\omega)|^2 +  |V_N(\omega)|^2\rangle 
   \\ &&+ 
  \lambda_1  \overline{H}_S(\omega) \overline{\Phi}_S(\omega) + \lambda_2   \overline{H}_D(\omega)  \overline{\Phi}_D (\omega)=0, \nonumber 
\end{eqnarray} with formal solution 

\begin{equation}
  \label{eq:optimalG}
  G(\omega) =  \frac{ \lambda_1  \overline{H}_S(\omega) \overline{\Phi}_S(\omega) + \lambda_2   \overline{H}_D(\omega)  \overline{\Phi}_D (\omega)}{\langle|V_T(\omega)|^2\rangle  + \langle |V_N(\omega)|^2\rangle}. 
\end{equation}

The second and third equations from Eq. \eqref{eq:partials} are the
same as Eq. \eqref{eq:Orthogonality} and Eq. \eqref{eq:Calibration}
above. The problem is then reduced to finding $\lambda_1$, $\lambda_2$
which (through the above), make $G(\omega)$ satisfy the two
constraints.

Substituting Eq. \eqref{eq:optimalG} into Eq. \eqref{eq:Orthogonality} and
Eq. \eqref{eq:Calibration}, we find

\begin{equation}
  \label{eq:Olambda}
  O_1 \lambda_1+O_2 \lambda_2=0,
\end{equation} and
\begin{equation}
  \label{eq:Clambda}
  C_1 \lambda_1 + C_2 \lambda_2 = \Phi_0.
\end{equation} where

\begin{equation}
  \label{eq:O1}
  O_1\equiv \int_{-\infty}^{\infty} \frac{|\overline{H}_S(\omega)|^2 |\overline{\Phi}_S(\omega)|^2}{ \langle |V_T(\omega)|^2\rangle  + \langle |V_N(\omega)|^2\rangle}d\omega,
\end{equation}

\begin{equation}
  \label{eq:O2}
  O_2\equiv \int_{-\infty}^{\infty} \frac{\overline{H}^*_D(\omega) \overline{\Phi}^*_S(\omega)\cdot\overline{H}_S(\omega) \overline{\Phi}_S(\omega)}{ \langle |V_T(\omega)|^2\rangle  + \langle |V_N(\omega)|^2\rangle} d\omega,
\end{equation}

\begin{equation}
  \label{eq:C1}
  C_1\equiv \int_{-\infty}^{\infty} \frac{\overline{H}^*_S(\omega) \overline{\Phi}^*_S(\omega)\cdot\overline{H}_D(\omega) \overline{\Phi}_S(\omega)}{ \langle |V_T(\omega)|^2\rangle  + \langle |V_N(\omega)|^2\rangle}d\omega,
\end{equation}

\begin{equation}
  \label{eq:C2}
  C_2\equiv \int_{-\infty}^{\infty} \frac{|\overline{H}_D(\omega)|^2 |\overline{\Phi}_S(\omega)|^2}{ \langle |V_T(\omega)|^2\rangle  + \langle |V_N(\omega)|^2\rangle}d\omega,
\end{equation}

with $\Phi_0\equiv \overline{\Phi}_S({\omega=0})$. The solution to the set of Eqs. \eqref{eq:Olambda} and
\eqref{eq:Clambda} is then given by
\begin{equation}
  \label{eq:lamba12}
  \lambda_1=\frac{\Phi_0 O_2}{C_1 O_2-C_2O_1}, \quad \lambda_2= \frac{\Phi_0 O_1}{C_2O_1 -C_1O_2}. 
\end{equation}

It should be noted that quantum noise is not explicitly considered in
the model. Rather, it is implicitly present in $\phi_H, \phi_V$ which
may differ from their average values $\bar{\phi}_H,\bar{\phi}_V$ due
to quantum noise. Note that the point of this measurement design is to
optimize the measurement of $\int_\mathcal{T}
\phi_H(t)-\phi_V(t)dt,$ including the quantum noise in that
variable. For this reason, it is sufficient to describe, and minimize,
the other contributions.

\section{Wiener filter estimator}
\label{sec:wienerest}

The Wiener filter estimator $\SWie$ can be derived from the frequency
domain Wiener filter output $\hat{X}(\omega)$ \cite{Vaseghi2000} define as

\begin{equation}
  \label{eq:wr1}
  \hat{X}(\omega)\equiv W(\omega) V(\omega),
\end{equation} where $W(\omega)$ and $V(\omega)$ are the Wiener filter and the electronic output in frequency domain, respectively. 

We define $W'(\omega)\equiv W^*(\omega)$ and $w'(t)\equiv w^*(t)$ and
make use of the inner product of the Parseval's theorem, see
Eq. \eqref{eq:innerparseval}.

\begin{equation}
  \label{eq:wr2}
  \int_{-\infty}^{\infty}W'(\omega) V_\mathrm{out}(\omega)d\omega=\int_{-\infty}^{\infty}w'(t) v_\mathrm{out}(t)dt.
\end{equation}

Then the Wiener filter estimator $\SWie$ is
$\int_{-\infty}^{\infty}w'(t) v_\mathrm{out}(t)dt$ corresponding to
Eq. \eqref{eq:raw0} for $\gamma(t)=w'(t)$.

The Wiener filter $W(\omega)$ is
\begin{equation}
  \label{eq:wr3}
  W(\omega)=\frac{\langle | V_\mathrm{ideal}^*(\omega) V_\mathrm{out}(\omega)|\rangle}{\langle | V_\mathrm{out}(\omega)|^2\rangle}.
\end{equation}

In order to compute the Wiener filter it is necessary to construct the
ideal signal $V_\mathrm{ideal}(\omega)$, a signal without all noise
contributions.

\acknowledgments

We thank F. Wolfgramm, F. Mart\'{\i}n Ciurana, J. P. Torres,
F. Beduini and J. Zieli\'nska for helpful discussions.  This work was
supported by the European Research Council project ``AQUMET'', the
Spanish MINECO project ``MAGO'' (Ref. FIS2011-23520), and by
Fundaci\'{o} Privada CELLEX Barcelona.  Y. A. de I. A. was supported
by the scholarship BES-2009-017461, under project FIS2007-60179.

\end{document}